\documentclass[preprint,showpacs,preprintnumbers,amsmath,amssymb,showkeys]{revtex4}

\usepackage{graphicx}

\newcommand{\dibuno}{\centering\includegraphics[width=9.5cm]} 
\newcommand{\beq}{\begin{equation}} \newcommand{\eeq}{\end{equation}}
\newcommand{\beqa}{\begin{eqnarray}}
\newcommand{\eeqa}{\end{eqnarray}}
\newcommand{\bneg}{\begin{bfseries}}
\newcommand{\eneg}{\end{bfseries}}
\newcommand{\biz}{\begin{flushleft}}
\newcommand{\eiz}{\end{flushleft}} 

\begin{document}

\title{Microscopic mass estimations}

\author{Jorge G.~Hirsch} \affiliation{ {Instituto de Ciencias
    Nucleares, Universidad Nacional Aut\'{o}noma de M\'{e}xico, 04510
    M\'{e}xico, D.F., Mexico}}

\author{Joel Mendoza-Temis} \affiliation{ {Instituto de Ciencias
    Nucleares, Universidad Nacional Aut\'{o}noma de M\'{e}xico, 04510
    M\'{e}xico, D.F., Mexico}}


\begin{abstract}
The quest to build a mass formula which have in it the most relevant microscopic contributions is analyzed.
Inspired in the successful Duflo-Zuker mass description, the challenges to describe the shell closures in a more transparent but equally powerful formalism are discussed.
\end{abstract}
\maketitle

\section{Introduction}

Understanding nuclear masses provides a test of our basic knowledge of nuclear structure.
Its accurate knowledge is relevant for the description of various nuclear and astrophysical processes \cite{Rol88}, 
Though great progress has been made in the challenging task of measuring the mass of short-lived nuclei
which are far from the region of stable, naturally occurring isotopes, theory is needed to predict their properties and guide experiments that search, for example, for regions of increased stability~\cite{Bla06}.

Advances in the calculation of atomic masses have been hampered by the absence of a full theory of the nuclear
interaction and by the difficulties inherent to quantum many-body calculations.  There has been much work in developing mass formulas with both microscopic and macroscopic input, on one side, and on the derivation of masses in a fully microscopic framework, on the other \cite{Lunn03}. The most successful approaches are the Finite Range Droplet Model (FRDM) \cite{Mol95}, the Skyrme and Gogny Hartee Fock Bogolyubov (HFB) \cite{Gor01}, and the Duflo-Zuker mass formula \cite{Duf94,Zuk94,Duf95}. They allow for the calculation of masses, charge radii, deformations, and in some cases also fission barriers.
They all contain a macroscopic sector which resembles the Liquid Drop Mass (LDM) formula, including volume and surface terms, the Coulomb interaction between protons, Wigner and symmetry terms, linear and quadratic in the neutron excess $N-Z$, and pairing. They also include deformation effects in different ways. HFB calculations have succeeded  in going through the 1 MeV barrier, which until very recently seemed unsurmountable, and have achieved RMS deviations smaller than .6 MeV \cite{Gor09}, competitive with the published fits obtained with the FRDM, which are also being improved.  On the other hand, the astonishing RMS around 0.3 MeV obtained by the DZ model 15 years ago sets the standard and provides the most robust predictions \cite{Lunn03,Men08}.

The Liquid Drop Mass (LDM) formula captures the macroscopic features of the mass dependence on the number of neutrons $N$, of protons $Z$, and on its mass numbers $A=N+Z$. Shell effects refer to the differences between the experimental binding energies \cite{AME03} and the LDM predictions.  Based in the remarkable success of the Duflo-Zuker (DZ) mass formula \cite{Duf94,Zuk94,Duf95}, the nuclear monopole Hamiltonian \cite{Duf96} allowed the description of the energy spectra of particle and hole states and particle-hole gaps on double magic cores \cite{DZ99}. Alternatively, there have been various attempts to describe the shell effects through a simple inclusion of linear and quadratic functions of the number of valence nucleons.  They improve the description reducing the root mean square deviation (RMS) by half \cite{Die07,JMT}. When 2-, 3- and 4-body terms are included, following the simplest DZ approach, the shell corrections for nuclear masses and radii can be calculated \cite{Die09}. 

The DZ mass model provides an attractive combination of simplicity and microscopic components. Since its initial formulation \cite{Duf94,Zuk94,Duf95}, there have been efforts to communicate its philosophy \cite{Cau05,Zuk08}. A detailed description of the simpler DZ10 model, which contains the basic ingredients, is available \cite{Men09}.  In the present contribution we present some general ideas for building a microscopic mass formula and speculate about the way in which they could be implemented.

\section{The macroscopic sector}

The macroscopic liquid drop model  employed in the present analysis is a slightly refined version of the Liquid Drop Model. It contains a {\em volume} term, proportional to $A$,  and a {\em surface} term, proportional to $A^{2/3}$, a {\em Coulomb} term
\beq
e_{Coul}=\frac{Z(Z-1)+0.76[Z(Z-1)^{2/3}]}{r_{c}},
\label{coul}
\eeq
which includes the charge radius $$ r_{c}=A^{1/3} \left[1-\left(\frac{T}{A}\right)^{2}\right],$$ instead of the mass
  radius $$r_{a}=A^{1/3} \left[1-\left(\frac{T}{A}\right)^{2}\right]^{2}.$$ 
They both include the dependence of the nuclear radii with the isospin $T=|N-Z|/2$.

The {\em asymmetry} term depends on the total isospin $T$ \beq
e_{Asym}=\frac{4T(T+1)}{A^{2/3}r_{a}},
\label{asym}
\eeq

The {\em surface asymmetry} is
 \beq
e_{SAsym}= \frac{4T(T+1)}{A^{2/3}r^{2}_{a}}-\frac{4T(T-\frac{1}{2})}{A
  r^{4}_{a}},
\label{wig}
\eeq

The {\em Pairing} term includes corrections of order $\frac{2T}{A}$ and scales as $1/r_{a} \approx 1/A^{1/3}$. It has different functional forms depending on whether N and Z are even or odd, listed in Table \ref{epair}.

\begin{table}
\begin{tabular}{cccc}
N & Z & ~~~~~~~~~~~~~~~~~~~~~~~~~~& $e_{Pair}$ \\ \hline
even & even & & $(2 - \frac{2T}{A})/{r_{a}} $\\
even & odd & $N>Z$ &$(1 - \frac{2T}{A})/{r_{a}}$ \\
odd & even & $N>Z$ &$1 / r_{a}$\\
even & odd & $N<Z$ &$1 / r_{a}$\\
odd & even & $N<Z$ &$(1 - \frac{2T}{A})/{r_{a}}$\\
odd & odd & & $\frac{2T}{Ar_{a}}$\\
\end{tabular}
\caption{The different expressions employed for the pairing contribution $e_{Pair}$ are listed in the fourth column. The six cases are classified with the parity of N and Z, listed in the first two columns, and with N larger or smaller than Z, third column.}
\label{epair}
\end{table}

\bigskip

With this six terms the macroscopic binding energy
is built
\beqa BE_{LDM} = a_{1} \,A -a_{2} A^{2/3} +
a_{3} \, e_{Coul} \nonumber \\ - a_{4} \, e_{Asym} + a_{5} \, e_{SAsym}
+ a_{6} \, e_{Pair} .
\label{macro}
\eeqa

The coefficients $a_i$ have units of MeV. They were selected to minimize the root mean square
deviation (RMS) when the predicted binding energies $BE_{\rm th}(N,Z)$
are compared with the experimental ones $BE_{\rm exp}(N,Z)$, reported
in AME03, modified so as to include more realistically the electron
binding energies as explained in Appendix A of Lunney, Pearson and
Thibault~\cite{Lunn03}. 

\begin{equation}
{\rm RMS}=\left\{\frac{{\sum\left[BE_{\rm exp}(N,Z)-BE_{\rm
        th}(N,Z)\right]^2}}{N_{nucl}}\right\}^{1/2}.
\end{equation}
$N_{nucl}$ is the number of nuclei for $N,Z\geq8$. The minimization
procedure uses the routine Minuit \cite{Minuit}.

\begin{figure}[!ht]
\includegraphics[width=19.0cm,angle=-90]{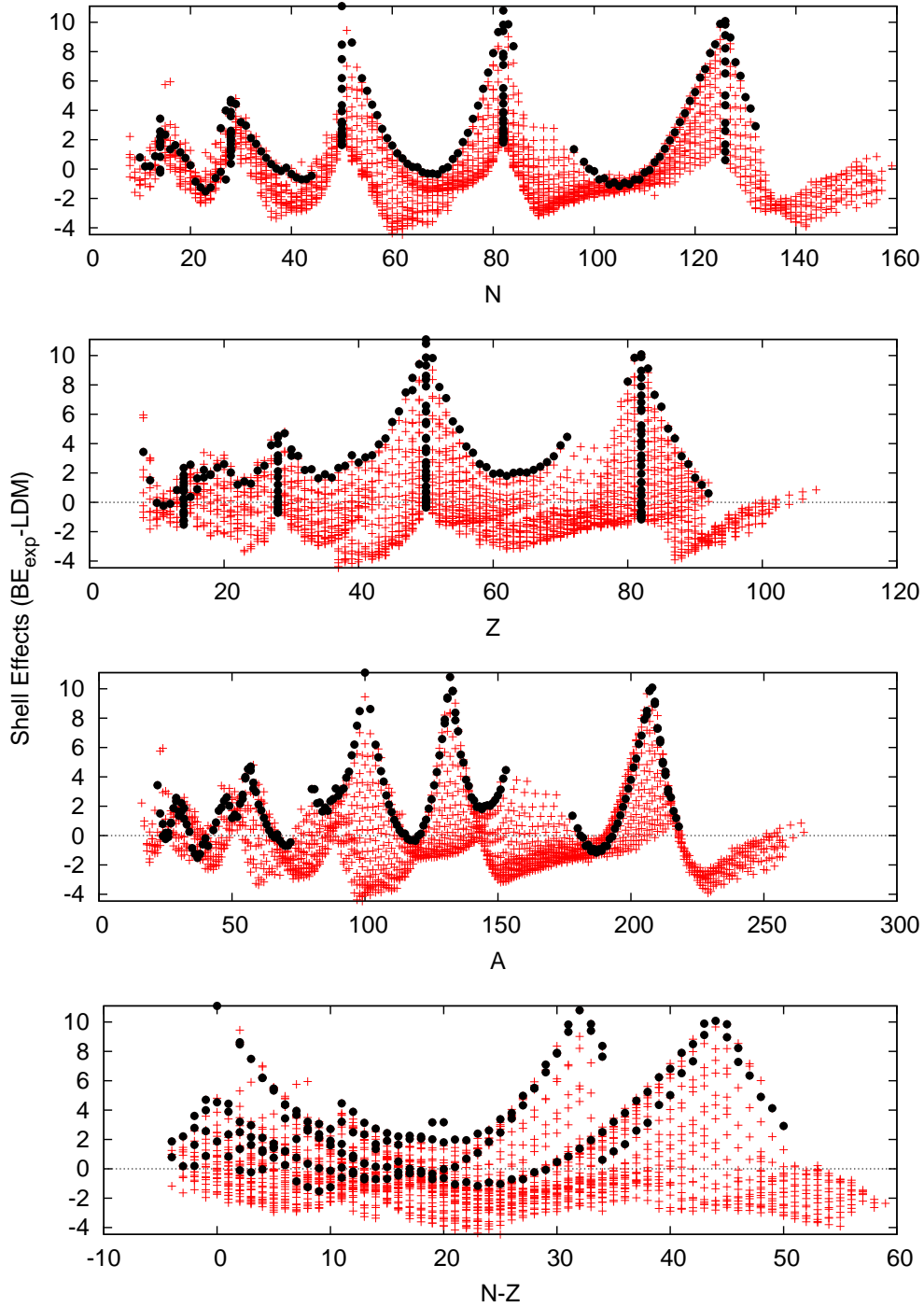} 
\caption{Shell effects, in MeV, for the Liquid Drop Model, for the whole set of nuclei in AME03.
and for the semimagic nuclei (heavy black dots). }
\label{dif_gota}
\end{figure}

The shell effects, i.e. the differences between the binding energies evaluated with the macroscopic LDM, Eq, (\ref{macro}), and the measured ones taken form AME03, are plotted in Fig, \ref{dif_gota} as functions of $N, Z, A$ and $N-Z$. They display information for the 2149 nuclei with N,Z $\geq$ 8.
To visualize the  binding energies when only protons or neutron shells are open, we present also results obtained employing only the masses of the 185 nuclei in AME03 which are semimagic, i.e. those with $N=14, 28, 50, 82, 126$ or $Z=14, 28, 50, 82$. The extra binding associated with the shell closures are clearly seen. These effects are the target to be described by the microscopic terms. Notice that the closed shell peaks are connected by smooth parabolas, evident in the plot for the semimagic, while the bottom regions are flat, a deformation effect taken explicitly into account in the DZ construction \cite{Duf94,Zuk94}. We can also observe that there is very little dependence on the isospin. The fitted  coefficients are presented in the second column of Table \ref{ajuste}. The RMS of the fit is 2.41 MeV.

\section{The shell closures}

One special aspect of the DZ approach is its ability to build the volume and surface terms starting from the monopole interaction.  The basic building blocks of the {\em master} terms are weighted sums over the occupied orbitals 
\beq
e_{1\nu}=\sum^{p_{\nu F}}_{p_{\nu}}\frac{n_{\nu}}{\sqrt{D_{\nu}}},~~~
e_{1\pi}=\sum^{p_{\pi F} }_{p_{\pi}}\frac{n_{\pi}}{\sqrt{D_{\pi}}},~~~
\label{e1}
\eeq
where the sums are defined for neutrons ($\nu$) and protons ($\pi$), and they run over all occupied 
$p_{\nu}, p_{\pi}$ orbitals up to the Fermi level $p_{\nu F}, p_{\nu F} $ with occupations $ n_{\nu}, n_{\pi}$ and degeneracies $D_{\nu}, D_{\pi}$. 

They are employed to construct the master term:
\beq
MA = {\frac {1} {2 r_a}}  \left( e_{1_\pi} + e_{1_\nu} \right)^2, 
\label{ma}
\eeq

This expression represents the dominant monopole component of the two-body interaction. The weights $ \frac{1}{\sqrt{D}}$ were fitted from the dominant term of realistic interactions \cite{Cau05,Zuk08,Men09}. 

Performing the sums for closed harmonic oscillator shells $p$, with degeneracies  $D_{p}=(p+1)(p+2)$, it is not difficult to obtain the scaling of MA, which is proportional to the mass number A. In this way the volume term is constructed microscopically in DZ \cite{Duf94,Zuk94,Men09}. Its precise asymptotic behavior is
\beq
MA_a = {\frac {3^{4/3}} {8 A^{1/3}} } \left( N^{2/3} + Z^{2/3} \right)^2.
\eeq

\begin{figure}[!htbp]
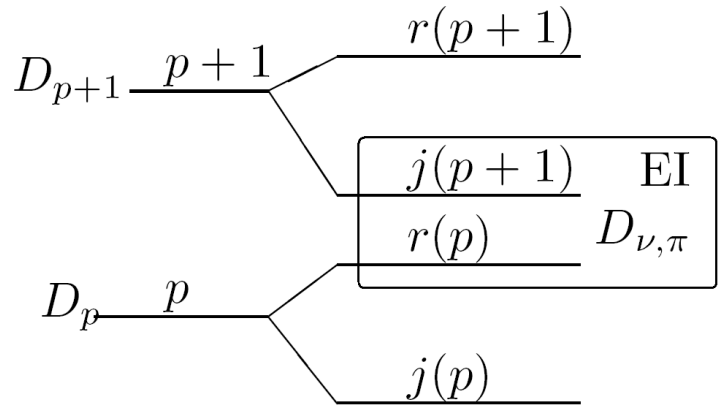

		\dibuno{CAPASEI.eps}
    \caption{Harmonic oscillator shells and extruder - intruder (EI) subshells. }
    \label{fig:EI}
\end{figure}

The difficulty with this approach is that the shell closures are the ones associated with the 3D harmonic oscillator (HO), and the mass model has to include other contributions which displace the closures form the HO to the extruder-intruder (EI) scheme shown in Fig. \ref{fig:EI}. 

The strategy which we want to explore here is based in the fact that expressions (\ref{e1}) and (\ref{ma}) can be redefined, with the index   $p_{\nu},p_{\pi}$ now referring to the EI major shell with degeneracies $D_{\nu,\pi}=D_{p_{\nu,\pi}}+2$. The fits of the monopole part of the two-body interaction cannot distinguish between the above mentioned one and this proposal \cite{Zuk09}. While it can be argued that the harmonic oscillator basis is the simplest and natural one to be employed in the microscopic description of the atomic nuclei, in what follows we show that the present proposal introduce in a very simple way the shell closures  of the Extruder-Intruder single particle scheme, at $N=14, 28, 50, 82, 126$ and at $Z=14, 28, 50, 82$.

We have explored two different ways to include the shell effects. In the first one we add one term to the binding energy, proportional to $MA - MA_a$. Its inclusion allows for a fit with an RMS of 1.46 MeV, a significant reduction for a single term. The fitted coefficients are listed in the third column of Table \ref{ajuste}.

\begin{figure}[!ht]
\includegraphics[width=21cm,angle=-90]{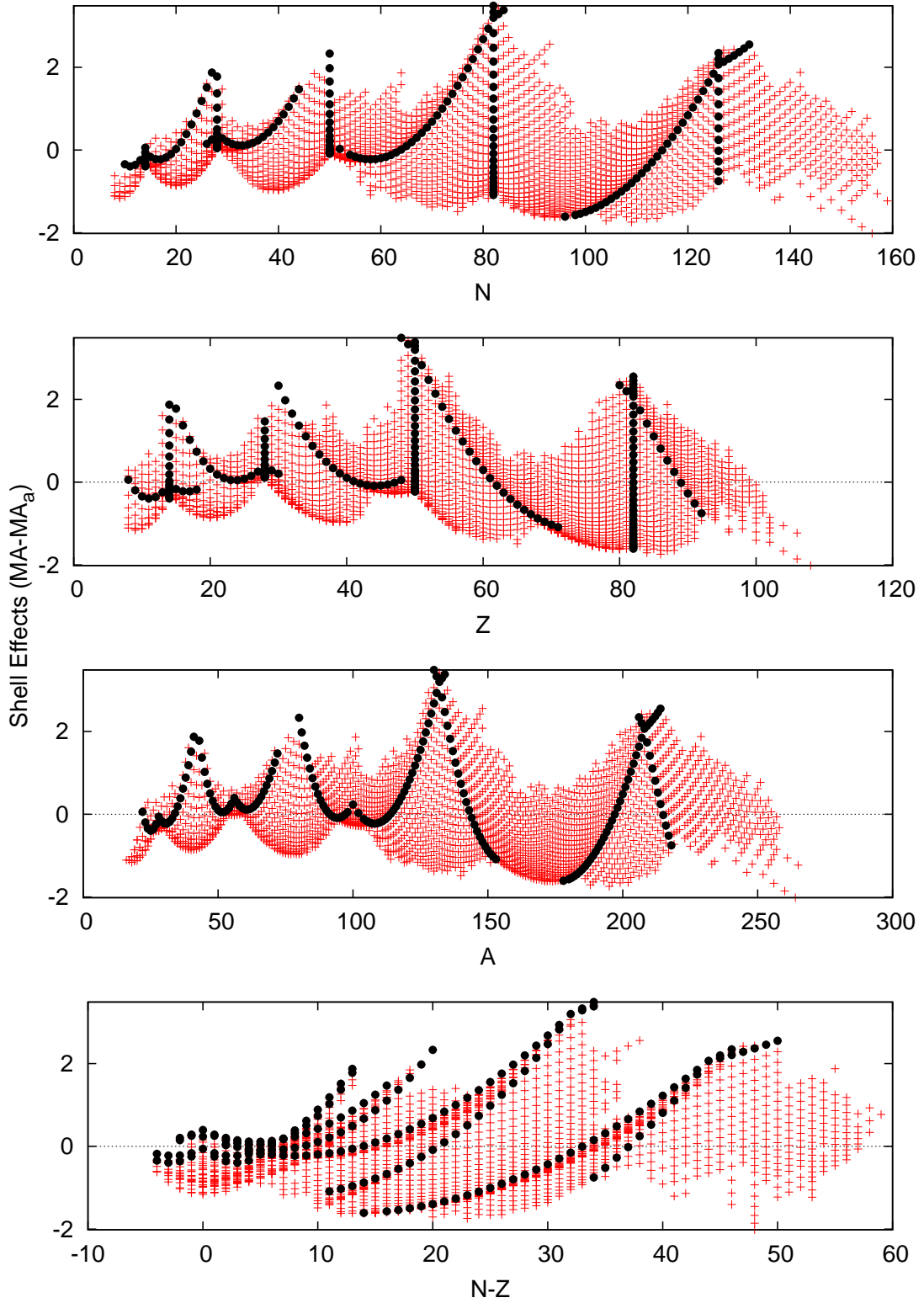} 
\caption{Shell effects, in MeV, for MA -  MA$_{a}$, for the whole set of nuclei in AME03 and for the semimagic nuclei (heavy black dots).}
\label{dif_ma}
\end{figure}

Fig. \ref{dif_ma} show $MA -  MA_a$ as function of N, Z, A and N-Z for the whole set of nuclei included in AME03 and for the semimagic nuclei. The peaks associated with the shell closures are clearly seen, but the parabolas are now partially truncated. In between these peaks, new parabolas have emerged, associated with open-shell nuclei.

\begin{table} [!h]
\centering
\caption{The fitted coefficients, in MeV, with their associated mean and RMS errors.}
\label{ajuste}
\begin{tabular}{|cc|c|c|cc|c|}
\hline ~~~~~& term & \multicolumn{2}{|c|}{~~~~~~coefficient~~~~~~ } & ~~~~~ & term & coefficient\\ \hline 
$a_1 $ & $ A $   	& 15.69054  & 15.10827  &$a_1 $ & $ MA$ 	& 15.99083 \\
$a_2 $ & $A^{2/3}$      & 18.69451  & 18.35651  &$a_2 $ & $MA/r$ 	& 17.62158 \\
$a_3 $ & $e_{Coul}$ 	&  0.71455  &  0.68267  &$a_3 $ & $e_{Coul}$ 	&  0.62218 \\
$a_4 $ & $e_{Asym}$  	& 26.48670  & 26.70229  &$a_4 $ & $e_{Asym}$ 	& 23.74367 \\
$a_5 $ & $e_{SAsym}$   	& 33.40293  & 28.81958  &$a_5 $ & $e_{SAsym}$  	& 26.84906 \\
$a_6 $ & $e_{Pair}$  	&  7.27923  &  6.64311  &$a_6 $ & $e_{Pair}$	&  5.93898 \\ 
\hline 
$a_7 $ & $MA - MA_a$    &   	    &  5.18889  &  	&               & \\
\multicolumn{2}{|c|}{constant}&  0.03458  &  6.63799  & \multicolumn{2}{c|}{constant}&  7.18870 \\ 
\hline
\multicolumn{2}{|c|}{\bf RMS} &  2.41383  &  1.46740  &      &   	&  3.33530 \\
\hline
\end{tabular}  
\end{table}

\begin{figure}[!ht]
\includegraphics[width=21cm,angle=-90]{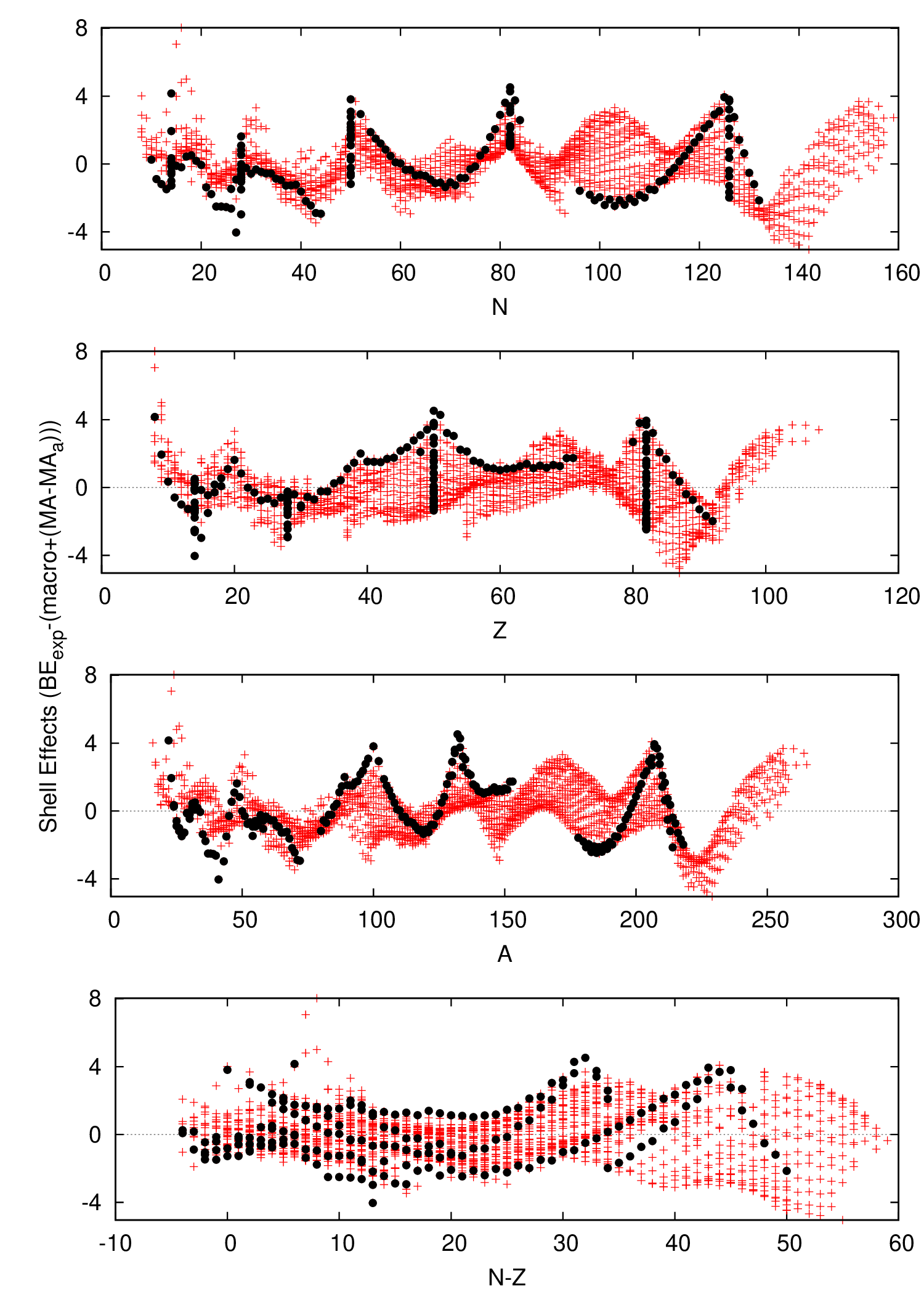} 
\caption{Shell effects, in MeV, for LDM  + MA -  MA$_a$ (RMS=1.46MeV), for the whole set of nuclei in AME03 and for the semimagic nuclei (heavy black dots).}
\label{dif_gota_ma}
\end{figure}

Fig. \ref{dif_gota_ma} displays the shell effects. i.e. the differences between the experimental binding energies and those obtained with the fit. The shell effects are still there, but diminished, with maximum differences around 4 MeV instead of 12 MeV for the pure LDM.

\begin{figure}[!ht]
\includegraphics[width=21cm,angle=-90]{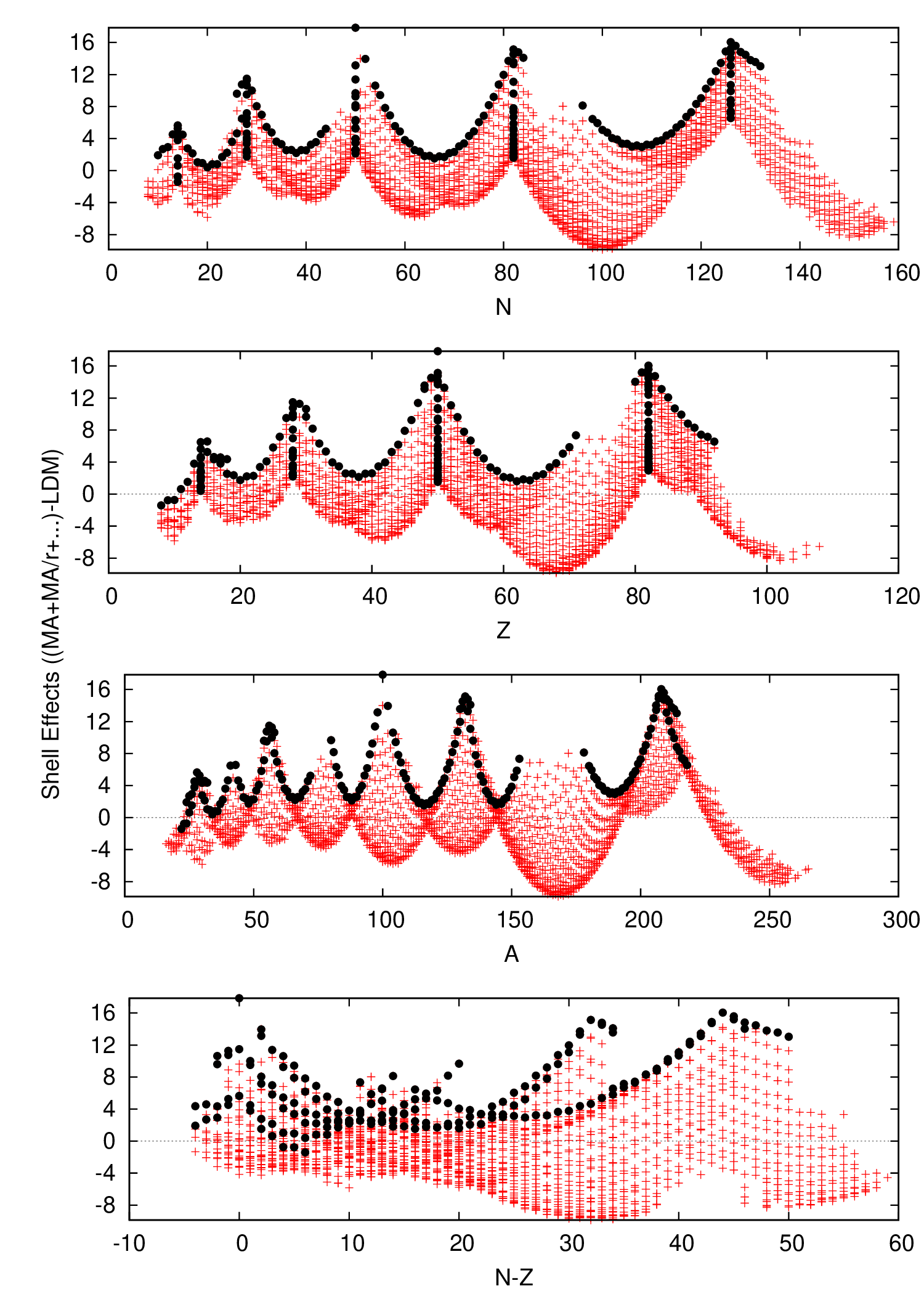} 
\caption{Differences between the binding energies obtained employing the master terms and the LDM, for the whole set of nuclei in AME03 and for the semimagic nuclei (heavy black dots), in MeV.}
\label{dif_master}
\end{figure}

The alternative approach is closer to DZ: to replace the volume and surface terms by $MA$ and $MA/r$.   
To get a glimpse of its significance, in Fig. \ref{dif_master} we have plotted the differences between the binding energies obtained employing the master terms and the LDM. It is clear that the master terms are building the shell effects. The image for the semimagic nuclei is particularly explicit.

\begin{figure}[!ht]
\includegraphics[width=21cm,angle=-90]{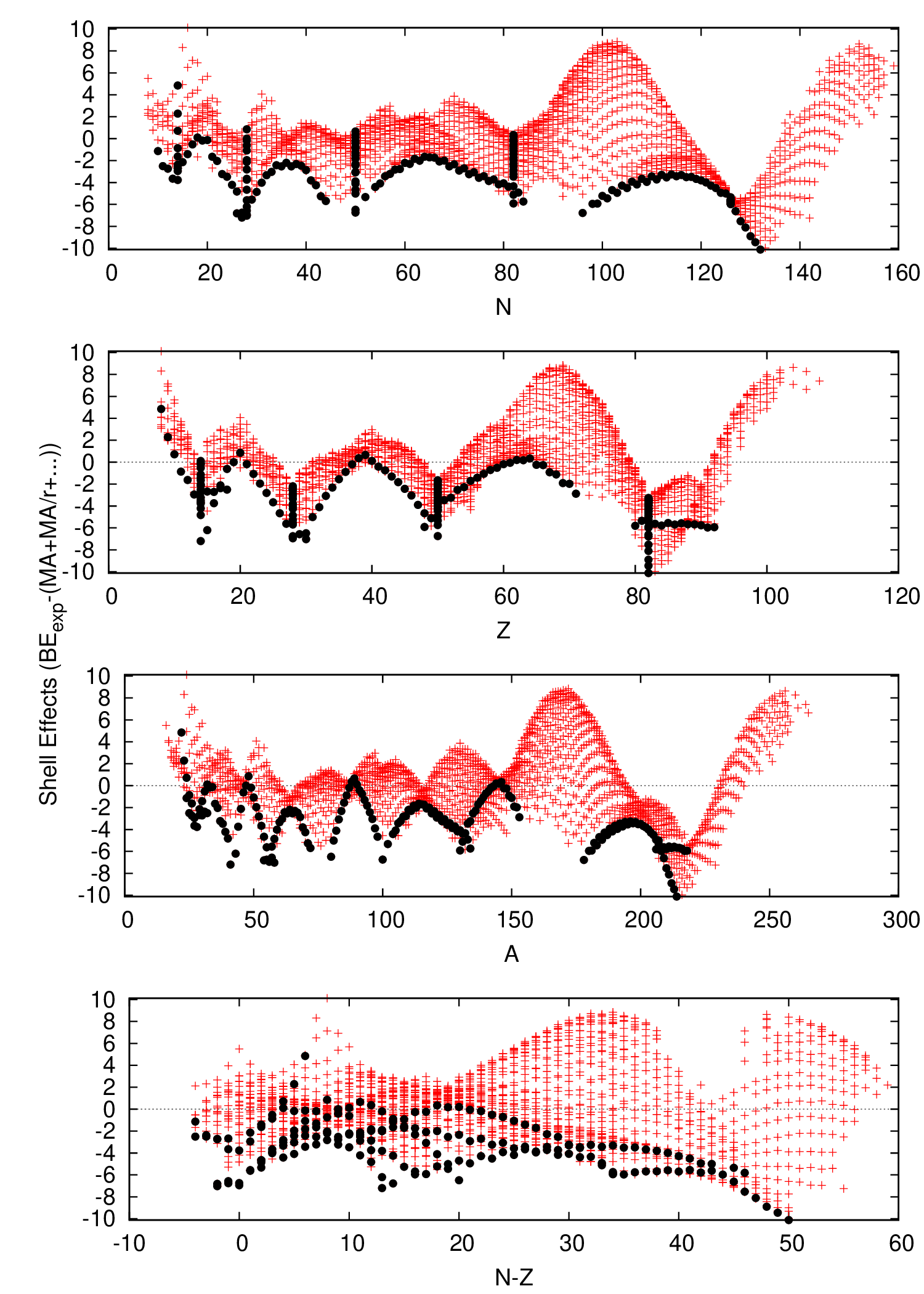} 
\caption{Shell effects, in MeV, for LDM with master terms, for the whole set of nuclei in AME03 and for the semimagic nuclei (heavy black dots).}
\label{dif_ma_mar}
\end{figure}

In Fig. \ref{dif_ma_mar} the shell effects for the LDM with the volume and surface terms replaced by the master terms are presented. The master terms are somehow overestimating the shell closures, which now appear with negative sign. The parabolas are there, and the RMS of the fit is 3.33 MeV, with the coefficients listed in the fifth column of Table \ref{ajuste}. 

Up to here we have explored the inclusion of the EI shell closures trough the use of the master terms with EI major shells. The next step would be to include two, three and four body terms, of the type employed in DZ.
All of them, being shell corrections, should scale as $r \approx A^{1/3}$, and vanish at shell closures. It means that {\bf only valence particles contribute to them}. At variance from the master terms, there are not sums here.
They can be built with operators which vanish at closed shells, i.e. with zero valence particles or valence holes. They have, for both protons and neutrons, the generic form

\beq
\Gamma^{1}_{ij} = \left( {\frac {m_i}{D_i}} - {\frac {m_i}{D_i}} \right) {\frac {D_i D_j} {D_i + D_j}}, ...
\eeq

The details can be found in Ref. \cite{Cau05,Men09}.

The other challenge is the inclusion of deformation effects. This was successfully performed in DZ exploring different occupation numbers in the intruder and normal parity subshells. Inspired in the Nilsson diagrams, deformed nuclei were assumed to have four protons and four neutrons removed form normal parity orbitals and occupying the next intruder orbital. From this point of view, the DZ model is a functional of the occupancies. One can visualize different models with increased complexity with DZ on one side and HFB on the other.   

\section{Conclusions}

In the present contribution we have explored the possibility of describing the shell closures employing master terms of the type proposed by Duflo and Zuker, but with major shells built in the EI scheme. It seems more promising to add a simple master term with its asymptotic behavior removed that to employ the full master terms instead of the volume and surface terms of the LDM. If this approach, when combined with the inclusion of two-, three- and four-body terms and allowing for different occupation numbers, can lead to an useful mass formula is indeed an open problem.   

\acknowledgments
 
The hospitality and exciting conversations with A.P. Zuker are sincerely acknowledged. JJMT thanks G. Mart\'\i nez Pinedo and K. Langanke for their support at GSI. This work was supported in part by Conacyt, M\'exico, and DGAPA, UNAM.

\end{document}